# A New Approach to Sonification of Astrophysical Data: The User Centred Design of SonoUno

Johanna Casado[1, 2, *], Beatriz García[1, 3], Poshak Gandhi[4], Wanda Díaz-Merced[5]

[1]Institute of Technologies in Detection and Astroparticles, National Atomic Energy Commission-National Council for Scientific and Technical Research-National University of Saint Martin, Mendoza-Buenos Aires, Argentina

[2]Bioengineering Institute, University of Mendoza, Mendoza, Argentina

[3]Mendoza Regional Faculty, National Technological University, Mendoza, Argentina

[4]School of Physics & Astronomy, University of Southampton, Southampton, UK

[5]European Gravitational Observatory, Pisa, Italy

**Email address:**
johanna.casado@iteda.cnea.gov.ar (J. Casado)
[*]Corresponding author

**Abstract:** Even when actual technologies present the potential to augment inclusion and the United Nations has been stablished the digital access to information as a human right, people with disabilities continuously faced barriers in their profession. In many cases, in sciences, the lack of accessible and user centred tools left behind researches with disabilities and not facilitate them to conduct front-line research by using their respective strengths. In this contribution, we discuss some hurdles and solutions relevant for using new technology for data analysis, analysing the barriers found by final users. A focus group session was conducted with nine people with and without visual impairment, using the tool sonoUno with one linear function and an astronomical data set downloaded from the Sloan Digital Sky Survey. As a result of the focus group study, incorporating data analysis using sonification, we conclude that functionally diverse people require tools to be autonomous, thereby enabling precision, certainty, effectiveness and efficiency in their work, resulting in enhanced equity. This can be achieved by pursuing a user-centred design approach as integral to software development, and by adapting resources according to the research objectives. Development of tools that empower people with wide-ranging abilities to not only access data using multi-sensorial techniques, but also address the current lack of inclusion, is sorely needed.

**Keywords:** Data Sonification, HCI, User Centred Design

## 1. Introduction

Digital technologies have substantial capacity for enhancing inclusion or, conversely, to further exclude people. Over the last two decades, a serious commitment to digital access targeting disability inclusion has been prioritised by international organisations. The United Nations (UN) enshrined, as a human right, the digital access to information (article 19 Universal Declaration of Human Rights-UDHR [1], and Convention on the Rights of Persons with Disabilities (CRPD) [2] annex g, article 2, article 4, article 9 section b, f, g, h, article 21, article 24). A booming growth of companies, initiatives, movements and efforts to equalise access has emerged recently, leading to the manifesto of Disability Interaction/eXperience, DIX [3]. Moreover, the UN went even further and, in article 24 of CRPD section e, states: "(e) Effective individualised support measures are provided in environments that maximise academic and social development, consistent with the goal of full inclusion." Here, 'individualised' refers to the body, the individual, and societal perspectives of each individual [4].

Though many initiatives exist, none has been able to improve the situation of people with disabilities in the educational field to be on par with the general populace.

Since education is a springboard to success in the labour market, this leaves the disabled in disadvantaged positions of remuneration. In our case, the sciences, people with disabilities (especially congenital disabilities) still perform sub-optimally in their profession, and in many cases, do not undertake managerial or mainstream work. This is not due to a lack of effort, but rather because there is a reticence to hire people with disabilities [5]. This is especially true for people with congenital disabilities: (i.e sensorial disabilities, orthopedically impaired, neuro-diverse and learning disabilities). The situation is completely different than late onset disabilities for people who already have job stability.

In terms of accessibility to digital interfaces, the paradigm of User Centred Design can benefit people with disabilities if it removes barriers, customising user experience (for example, when interacting with digital technologies, a congenitally blind user must be able to enjoy the freedom to choose navigational order; until now, their only option is sequential navigation). It is inexcusable that expensive machine learning algorithms are constantly monitoring our responses, adapting choices and reconfiguring the algorithm to fit our interests or needs and are scarcely used to adapt the accessibility interfaces to the user, making interfaces more user centred.

Current development/implementation of technological design and focus groups/usability evaluations, follow the International Standardisation Norm, WCAD, WC3, WAI-ARIA and the like. These are important norms where many specialists reach consensus. But if the recommendations are not implemented appropriately, this can instead serve, as mentioned above, to impair the voice of disabled users. Those approaches have an expectation of functioning for people with disabilities, with the danger of becoming an impersonal digitally embodied social arrangement (very far from the end user's real needs). For example, consider a usability evaluation of an interface that abled users can navigate without problem, but a disabled user who cannot is instead told "other people did manage to do it". According to the sociologist and UN human right collaborator [6], with such a comment the disabled user is effectively being told: "Your situation is fake". The faults of system design and architecture end up placing the onus of blame on the disabled instead: "The problem is that You are different!". By the same token, the disabled user may feel that the system is not taking her/him into consideration and that there are no choices but to give in. It is a violation of the Human Rights of Peoples with disabilities to disregard individual variations (CRPD [2]). This is evident in even cursory examinations of social thought, pervasive in the development of digital technologies which has directed itself to the notion that it is impossible to accommodate everyone, or that it is impossible to create a digital interface usable by everyone with user-centric interaction options.

The UDHR and the CHRPD mandate the development/implementation of a framework focusing on questions such as: "How do I enable the user to utilise this interface to perform optimally, without having to dramatically change her/his particular situation? What shall be implemented to maximise user experience, without submitting her/him to learning curves and transactions that may negatively hassle her/his self esteem and further exclude her/him"? The World Health organisation with its international classification of Functioning, Disability and Health, has launched the approach of treating the reality of each person as a variation and not as an abnormality.

The community of HCI specialists and computing sciences are changing the framework of focusing on "the impairment" of the individual (the supply side) to instead focus on an integrative framework that considers the individual impairment and its functioning (in context) and the capacities of the individual, an approach based on capacity not incapacity. Holloway, C. [3] proposes a mindset and framework of mainstreaming of disability in Human computing interaction and the digital world (in other words, to mainstream every interaction). In a very smart way, it calls for the integration of technologies to cater to people with disabilities. DIX does not call for people with disabilities to have choices or to decide. Disability is multidimensional, for that reason the disability interaction framework proposed in 2019 [3] lacks one question: how do I integrate technologies that involve sustainable reforms, not just short-term strategies? If the aim is to impact the social development of the individual, for disability to become mainstream, we need to create technologies that will not only accommodate individual abilities, but will also evolve as interaction styles evolve in time, and ultimately predict and adapt to interaction styles that may emerge in the future. How do we design technologies that permit people to function with what they already have/bring and just as they are?

The sonoUno project seeks to address the following topics:
1) Accessibility to scientific data, from the Earth or with instruments on board satellites (available in databases such as Simbad, NASA, ESA, amongst others).
2) Creation and maintenance of a human-computer interface suitable for the access, collection, sonification and analysis of astrophysical data.
3) Testing the efficiency and effectiveness of the resource in different cultural environments.
4) Development of a paradigm for training researchers and interested citizen scientists to start using these new techniques (the tool itself and sonification).

Taking in consideration that this work present a focus group analysis where sonoUno has been tested, the definitions of usability, efficiency and effectiveness have been adopted from the ISO 9241-11 standard [7]:
1) Usability: "extent to which a system, product or service can be used by specified users to achieve specified goals with effectiveness, efficiency and satisfaction in a specified context of use".
2) Efficiency: "resources used in relation to the results achieved".
3) Effectiveness: "accuracy and completeness with which users achieve specified goals".

## 2. Related Works

Visualisation of a large amount of information can be challenging, because of the resolution and limited attention focus of the human eye. This impact the dynamic range of visualisation. The usage of multisensory interfaces for data "visualisation" in visually impaired people has been developed in [8, 9], and even includes concepts verging on the artistic in [10, 11].

In recent years, the number of sonification tools that allow specific data to be audified has grown. Some with the capability to work with astronomy-specific datasets are Sonification Sandbox [12], MathTrax (https://prime.jsc.nasa.gov/mathtrax/), xSonify [13], Sonipy [14] (https://github.com/lockepatton/sonipy) and StarSound [15]. Most of these present graphic user interfaces, except for Sonipy that can be used in the 'bash' window shell by importing libraries through the 'python' environment. Sonification Sandbox presents several notebooks that demonstrate the different software functionalities; it uses MIDI libraries for a variety of sound instruments to sonify the data sets. MathTrax has an educational focus, with the option to plot and sonify basic mathematical formulae, and with different notebooks for each group of functionalities as in Sonification Sandbox. In contrast, xSonify is centred on sonifying astronomical data, presenting an octave bridge to allow data processing (this feature is not available now). Finally, StarSound contains a sound synthesiser that can be fine-tuned by the user, presenting an option of adjusting it with a text field, especially for visually impaired end users.

## 3. SonoUno Development

### *3.1. Theoretical Framework for a User Centred Interface*

The SonoUno software presents a user-centred front-end design from the start. The first beta version, before the prototype, was a mock up design comprising the first group of functionalities. These were chosen over the course of several meetings held at the International Astronomical Union Office for Astronomy Development (OAD), in South Africa (Figure 1). This design was inspired by a theoretical framework centred on accessibility for blind people. Considerations around this framework led the team to write some recommendations for accessible human computer interfaces, published in [16].

From these recommendations and utilisation of the authors' respective expertise, some casual parameter rules were defined and detailed in [17]. These include: (1) there must be a linear relationship between the functionalities; (2) the categorization of functionalities should be unique and no overlap with each other; (3) no existence of hidden functionalities; (4) the interface display should be simple.

In addition, Bahr and Ford have shared the feedback that "users (participants) considered pop-ups annoying and frustrating and did not enjoy pop-ups" [18] (p. 781). Massengale, L. R. and Vasquez III, E. [19], during an analysis of accessibility in online courses, describe some challenges that users faced and one of these is "Content opens in pop-up windows". It is notable that there are not studies with end users exclusively about pop-up windows, but it is known that there are algorithms that automatically detect and avoid pop-up windows in digital interfaces.

Taking into account all these guidelines the sonoUno framework was designed from the beginning to be orderly and intuitive (in terms of arrangement and usage of functionalities) (Figure 2). A selected number of functionalities were placed into panels that the user can expand and contract according to their desire (Figure 3). This facilitates the user to decide upon the number of elements accessible at any one time on the interface.

### *3.2. SonoUno Description*

The task of coding sonoUno was divided up into work packages. Each programmer took charge of one of the following work packages: (1) sound package, (2) graphic user interface (GUI) package, and (3) input/output (I/O) package, resulting in three features per iteration (Figure 4). This was documented once the code of the first prototype had been written and tested.

Python was chosen as the programming language of choice. Due to its large community of developers and rapid adoption, it was sought to sustain the maintenance of the software over the long term, both with the update of its scripts and the libraries it uses. SonoUno makes use of the following libraries: wxPython for GUI, matplotlib and numpy for data processing, pygame for sound synthesis, and oct2py for the octave bridge.

The principal framework of the GUI (Figure 2) contains the basic elements to control the reproduction of data sonification, following the principles established in the previous section. Then, the end user can find additional features in the menu bar such as I/O functions (file menu), basic mathematical functions (data operation), sound and plot settings, as well as help. The menu item panels allow opening and closing of panels (Figure 3) that control the different features; this design permits sonoUno to have a lot of functionalities without the user having to deal with pop-up windows.

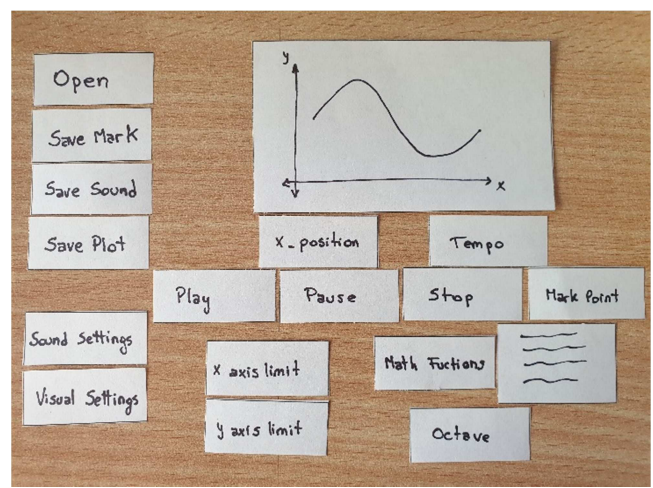

***Figure 1.*** *SonoUno's first mock-up, designed during meetings in IAU-OAD South Africa [2018] (https://www.astro4dev.org)*

45  Johanna Casado *et al.*:  A New Approach to Sonification of Astrophysical Data: The User Centred Design of SonoUno

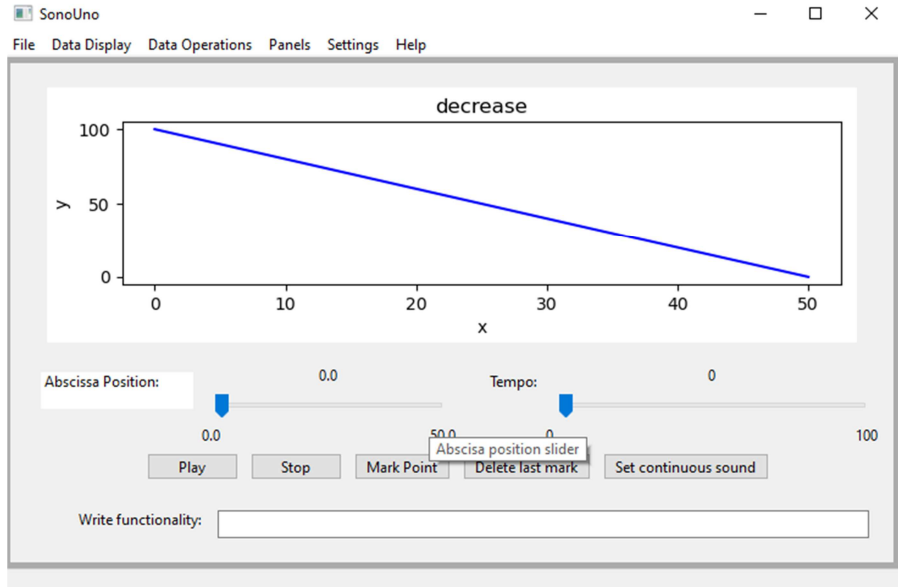

*Figure 2. The main GUI of sonoUno, presenting a menu bar, an example plot with an inverse function, and several sound reproduction and data position buttons.*

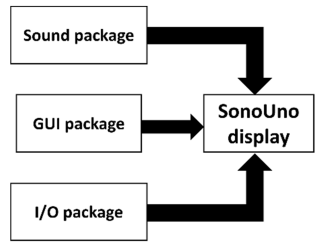

*Figure 3. SonoUno actual display with the principal panels enabled: 1-Input/Output panel (equivalent to File in the Menu bar); 2-Settings panel; 3-Display panel; 4-Maths panel. An example data file has been loaded and displayed in panel 3, this dataset was provided by members of the REINFORCE project ((GA 872859) with the support of the EC Research Innovation Action under the H2020 Programme SwafS-2019-1 the REINFORCE).*

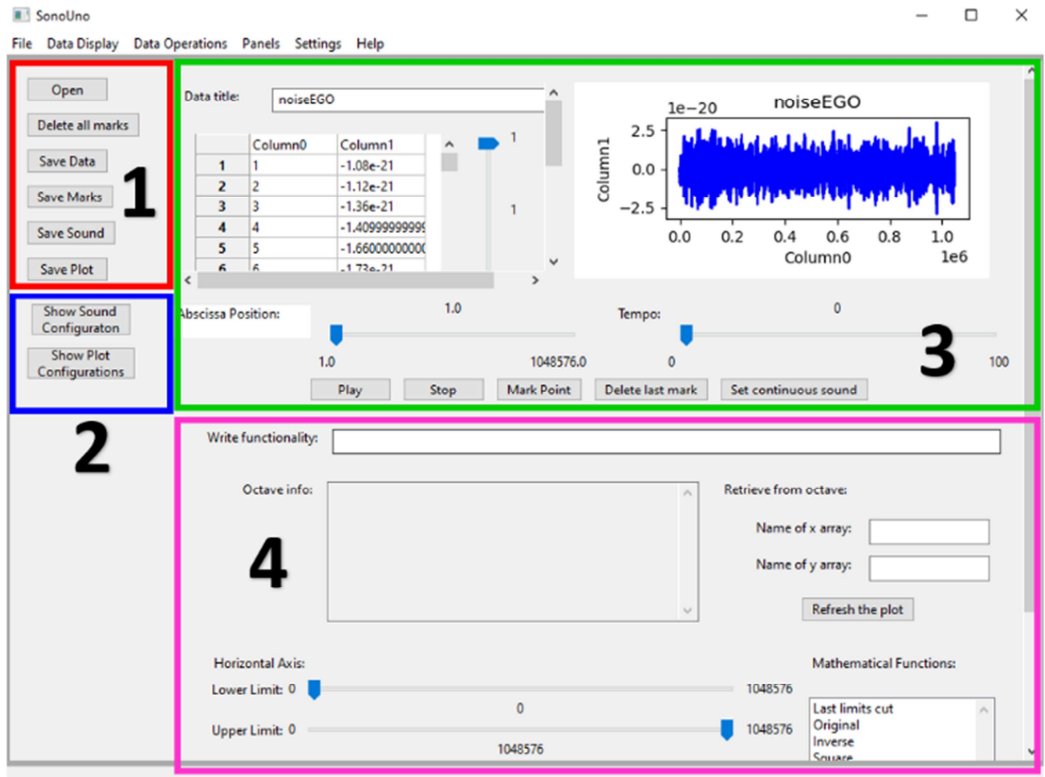

*Figure 4. The first three work packages (sound, GUI and I/O) that formed the foundation of the sonoUno display.*

## 4. Focus Group Analysis

### 4.1. Methodology

With the first version of the sonoUno software ready, we investigated its effectiveness and usability by people both with and without disabilities. This was done through a focus group conducted in the city of Southampton, UK, during April 2019. In addition, the focus group sought to consult about sound quality, and the opinion of the participants on this new multimodal approach proposal for the analysis of astrophysical data.

Consequently, the focus group was enriched with a series of tasks designed with three levels of complexity (low, medium and high). The sessions were divided into two spaces that were held on the same day, with an intermission break that allowed participants to assimilate and interact (talk between them, for example) on the first use of the software, and then use the tool with a set of astronomical data provided by the team of investigation. The first session was based on an introductory question, following which the tool was presented to the participants. They were then asked to complete the list of tasks, and the session finished with questions on effectiveness of the tool (two of them were on software usability, one on sound, and the last one on the possibility of analysing data with this tool). As for the second session, after a 15-minute break the first task was to test the program with a set of astronomical data downloaded from the Sloan Digital Sky Survey database (http://skyserver.sdss.org/dr12/en/tools/quicklook/summary.aspx?id=1237648720693755918), any list of task was provided, people could use the program as they desired; after that, the session continued with further questions: five focused on program usability, one on sound and two for their opinion on the multimodal approach to data and whether they would use it in their work (the questions and list of task are available at: https://drive.google.com/drive/folders/10ywUuI4hdnho_x7WJMzq11hUqe-D4F95?usp=sharing).

Nine people invited by email and personal contact agreed to take part in the work and were divided into four groups according to whether they work in astrophysics and their functional diversity:

1) Group A: 4 people who do not work with astrophysical data in their day to day (2 with low vision, 2 totally blind). Four of them use the screen reader with the software and 2 of them additionally used the magnifier glass, the tools are preinstalled on the computer.
2) Group B: 1 person not sensorially deprived, no astronomer as a profession, background in computing science.
3) Group C: 1 person with low vision with background in astronomy (professionally). This person used the screen reader in conjunction with the magnifier glass.
4) Group D: 3 people not sensorially deprived, professional background in astronomy (Age range: 22-27).

Due to time and availability constraints, the groups B and C could not be enlarged. It is extraordinarily difficult to gather a good sample of visually impaired people who work actively in the field of astrophysics (or any other field), especially during early onset, because there are currently no tools that allow them to carry out their work autonomously. This is one of the motivating factors for our research and one of the objectives of the tool under development. As for Group B comprising only one person, it was considered important to have at least one person without disabilities who did not work in astrophysics.

Recognizing the above potential limitations of the conformed groups may, it is considered that the groups are still sufficient to obtain first answers to our research questions, which are: 1) Is the software under development usable, effective and efficient?; 2) Does the package allow data analysis through sound?; 3) What is the opinion of the participants regarding multimodal analysis of astrophysical data?

All the focus group meetings were carried out in an analogous way and audio was recorded following participants' consent. The project also obtained ethics approval prior to start (ERGO II reference 48331). Once the focus group meetings finished, the audio recordings were transcribed for later analysis and destroyed after a period of one year.

### 4.2. Text Condensation Technique

The transcription analysis was carried out by two authors. First, the 'systematic text condensation' technique was used [20], through the necessary meetings indicated by the method. Thereafter, the analysis was continued separately until representative 'statements' about the transcripts were defined. A new meeting between the two authors was held to discuss and reach an agreement on the defined categories, and finally, working individually, the 'statement' defined in each category was finalised. The defined categories are:

#### 4.2.1. Memory Overload

Indicated by responses referring to the need to carry out a lot of transactions in order to reach their goal. Example feedback from the participants (identified anonymized codes) include: A4 during the main test "my instinct when trying to pause playback is instead to restart" (in the current sonoUno version, Play and Pause are assigned to different buttons); A2 following an instruction by the moderator ("You have to select an instrument") answer: "The sound font?" (here A2 was lost in the previous text control of instrument-select that only indicated that the software used a determined sound font; this unnecessary element diverted the attention of A2 and contributed to memory overload). In addition, declarations about lack of consistency also contribute to memory overload, referring to confusion about differences between graphic and auditory display: B expressed "when I saw the original graph, with the continuous line, I don't expect to hear individual notes". Moreover, there were positive remarks regarding this category: B expressed "everything was quite obvious" reinforcing the presence of consistency between other applications and sonoUno; in concordance, A4 declared "These are the same commands that are the default options in Notepad"; in the case of C, the expression during a functionality searching was "So, that's the third, yeap, ok".

*4.2.2. Information Needs*

Determined by participants' responses and direct indications on the need for the sonoUno prototype to keep them informed about each action in a multimodal way: B expressed "If you mark a point, what is the wavelength value?"; C1 also commented "And does it tell me what my selected range is?". Furthermore, the need for confirmation after an action arises (D2 commented "but I expect to see something here when I push 'mark point'"; the problem in this version was that the mark was placed behind the red bar and the user could only see it after the red line was moved). Furthermore, ensuring communication with the assistive technology and through the messages inside sonoUno was an important focus; in this sense, A4 expresses "it says 25 slider [...] what's slider mean?"; about assistive technology and the importance of providing the same tools to which people are used to, A2 highlights "[...] I need to be reminded, but that's because I prefer to have the information in braille besides me [...]"; in addition, A4 and A1 commented about the need of precision and feedback from sonoUno: A4 "[...] but by the time, if you listen, you hear the peak, by the time you get the shortcut key [...]" A1 "Yea, I was concerned [...]".

*4.2.3. Choice Needs*

Designated by participants' discourse where they highlighted that this tool offers to the users with disabilities the possibility to explore the data and control interactions from beginning to end: C expressed "[...] freedom as well, I felt like I was a lot more able to access the bitty data that I want to do"; A1 also commented "I think in the circumstances where you are working with people who have no, literally no vision, then it's quite valuable because it gives you access to something that you otherwise wouldn't have access to, so information that you wouldn't have access to". In contrast, visually-able participants declared the need for more features that allow working with the data and the graphic display: D2 said "I need some interactive things, really analysing on the command line"; B "[...] these bars, I would like to put specific numbers on" referring to the slider bars that allow changing the x axis limits on the plot. In addition, direct indications remarks on the need to control the interaction, and the need for precision of the action that the user performs, in conjunction with the possibility to adjust settings of the interface: for example, A2 asked during the second test after the break "So, is this a piano again?"; B pointed out "[...] take a lot of time to play the full graph"; B, D1, D2 and D3 expressed "[...] make the plot more interactive"; C declared "I think having the ability to change the instrument or the time is really important."

*4.2.4. Training Needs*

Determined by responses pointing out the need of help at the beginning in the use of the tool by visual impaired people: A3 and A2 exchanged "[...] if somebody could demonstrate how to open different folders, or the folder structure and how it looks, would it have been easier?" "Probably, yes"; also, A1 complemented "[...] but at this stage we certainly need people around", referring to the version of sonoUno tested. Moreover, participants' discourse also indicated the need of training about the general sonification technique: B mentioned "[...] I don't think that I could pick out the small trends"; D2 complemented with "it was not that easy to identify the things, just by sound"; A4 highlighted this with a question "I mean if you've got four pictures, how do you know that the hydrogen line of this picture is in the same position as the hydrogen line on the second one that you analyse, to know that is moving at the same speed or the same distance? 'cause at the moment you are doing it visually, I can't look at the picture, how does a blind person look at two audio pictures?"

*4.2.5. Social Aspects*

Indicated by participants' declarations where they charge in them the guilty of some software malfunction or highlight the misunderstanding that people have about disabilities: C expressed "[...] my only complaint would be the inability to increase the tempo even more than you can at the moment, because I'm impatient"; B commented "for blind people, how they can use the interface?"; D3 affirmed in response to any advantage of the software "for scientific reasons I don't know, for outreach reasons a lot". In addition, responses of people with disability pointed out the importance of user-centred developments that allow them to carry out scientific investigations: C indicated "eventually it would make the kind of research that I had to do even easier, much much easier than is possible now, allowing me to research things that I would found very difficult before".

*4.3. Reliability Analysis*

From this entire process, two summary category tables were constructed, where each analyst classified the statements by category and then agreement was measured using the kappa coefficient [21]. Lombard, M. et al. [22] emphasises the importance of content analysis and reliability; they also differentiate intercoder reliability as a measure of the degree to which different judges tend to agree. Hallgren, K. A. [23] mentions that kappa statistics are used for nominal variables and "is only suitable for fully-crossed design with exactly two coders" (p. 6). In addition, [24, 25] describe the kappa statistic method.

*Table 1.* *Agreement matrix of proportions obtained from the category tables of each coder [21].*

| Category number | | Coder A | | | | | |
|---|---|---|---|---|---|---|---|
| | | 1 | 2 | 3 | 4 | 5 | $P_{iB}$* |
| Coder B | 1 | 0.107 | 0 | 0 | 0 | 0 | 0.107 |
| | 2 | 0 | 0.178 | 0 | 0 | 0 | 0.178 |
| | 3 | 0 | 0.071 | 0.286 | 0.036 | 0.036 | 0.429 |
| | 4 | 0.036 | 0 | 0 | 0.107 | 0 | 0.143 |
| | 5 | 0 | 0 | 0 | 0 | 0.143 | 0.143 |
| | $P_{iA}$* | 0.143 | 0.249 | 0.286 | 0.143 | 0.179 | 1.00 |

* The $P_{iX}$ values are the marginal proportion that each coder assigns to each category, so $P_{iA}$ is the marginal proportion that coder A assigns to each category and $P_{iB}$ is the marginal proportion that coder B charges in each category.

Since we work with nominal data and the analysis was led by two coders, the kappa statistic was used as a measure of reliability. Table 1 shows an agreement matrix, from which the kappa coefficient is calculated: k = 0.768. Following the criteria of [21] the kappa (k) value illustrates the proportion of agreement after chance agreement is removed, and it is calculated as:

$$k = \frac{po - pc}{1 - pc} \quad (1)$$

where *po* is the proportion of units in which the judges agreed and *pc* is the proportion of units for which agreement is expected by change.

Employing (1) with the agreement matrix (Table 1), the kappa coefficient is 0.768. According to [25], a kappa equal to 1 represents perfect agreement and a value between .60- .79 is moderate. Considering that 0.76 is near the upper threshold, the coefficient indicates good agreement between coders, implying our results to be reliable.

### 4.4. Results

Following our previously defined categories, we are going to highlight some important findings during the text condensation technique.

About 'memory overload' category all participants recognize the functionality linearization that sonoUno presents, but A4 focuses on the fact that one menu was redundant. Therefore, four other participants were queried on this point and all agreed with group A. It was found that the separate play and pause buttons confuse the end user (these functionalities are on the same buttons in the actual release). Regarding consistency, B expects to hear continuous sound when the plot presents a continuous line (the current default sound is discrete), this presents an inconsistency between visual and audio display. On the other hand, this category demonstrates the importance of functionality linearization and consistency with other widely used applications, with all participants verifying that the generic functionalities are in their expected places and are very easy to find (for example save, open, play, stop).

Regarding 'information needs' category, all participants (100%) express the need for information and confirmation regarding their actions (some paraphrased comments: "Where am I? Where are the peaks? Am I pointing the peak correctly?"; "It would be good to know the selected range"; "It would be good to know the wavelength value of the marked point and the exact position on the abscissa"), in this sense, good communication with the user and with assistive technology must be assured. Specifically, an assistive technology dependency became evident during the focus group session – the lack of synchronisation between the prototype and the screen reader forces the visually impaired participants (55.5% of the total, 100% of visual impaired) to perform actions to get things running that are not part of the main task for the focus group, for example asking about "What does slider mean?". This is the case even when sonoUno uses the native accessibility configurations and it can be used with different screen readers, this highlights the need of better communications between the different applications and the assistive technology tools.

Concerning 'Choice needs' category, it emerges from the participants with disabilities (group A and C, 100% of participants in these groups), there are few tools that allow them to make selections and explore but, in general, with a huge number of transactions and high learning curve. Although, in the case of sonoUno during the focus group session participants express that this tool allows them to explore and choose between different options at will (see definition of this category in the previous section for textual cites) this is not common especially in the field of research. In this category, also the need for more sound settings emerged ("a variety of instruments that are noticeably different"; "to add continuous sound as well"); need for more accuracy ("I like to put specific numbers" - referring to cut sliders); need to retrace a few positions in order to hear recent sounds again; the possibility to change the panels sizes; and the chance to adjust font settings, among others.

Respecting 'training needs' category, during the focus group session with the different groups, the need for training for both the usage of prototype and the sonification technique arose. Sonification applied to analyse datasets is a new approach. Therefore, people express the need for guidance to gain confidence in the technique, together with guidelines to ensure that most people understand the same data in a cohesive manner. In the case of the prototype, it was evident that people during the second period were much more confident using the software; this showed us that the software presents a fast learning curve. Perhaps, a combination of training courses for sonification using sonoUno could be a good starting point, strengthening the technique along with making sonoUno more powerful and user-centred.

Relating to 'social aspects' category, it evidences the social problem that people with disabilities suffer and not disabled people: neglect and misunderstanding. During the session people with no disabilities considered the tool primarily as outreach software; none of them thought/expressed that people with disabilities can do science (D3 affirmed in response to any advantage of the software "for scientific reasons I don't know, for outreach reasons a lot"). On the other hand, people with disabilities expressed gratitude, and put on their own shoulders the blame for prototype mistakes (see this category definition in section 4.2). This is far from ideal, highlighting a huge misunderstanding regarding accessibility: every functional diverse person with the right tool can make whatever they desire if they have the will. The other possible enemy of this affirmation is the mindset of developers. If developers are not trained to consider the needs of functional diverse people in code development from the beginning, and listen, assess and integrate solutions that will adapt to their needs, the production of a usable and useful tool is very difficult.

Regarding the research questions expressed in section 4.1, in response to the first of these ("Is the software under development usable, effective and efficient?") It is evidenced that the code fulfils its main objective (display visually and sonify astronomical data). However, it presents a very simple interface, without the tools that an astronomer uses today in their profession, in addition to some complexity for the visually impaired.

About the second research question ("Does the package allow data analysis through sonification?"), all the participants managed to correctly identify (via audio) a decreasing function in the first data task set for them. In a second task, they correctly detected three peaks that correspond to three spectral emission lines. It is important to highlight that the participants without disabilities expressed that beyond being able to detect these patterns, at the moment they do not believe they can analyse data through sound, corresponding with [26]. But our findings do not close the door to sonification, the general discourse with our participants makes evident the lack of experience using sound and the novelty of the technique. All participants said that they would use the software in the future if they have the opportunity. Coming back to sonification, the participants with visual impairment also mentioned the importance of training for the sound technique. It is concluded from this that designing adequate training is of utmost importance to be able to start talking about analysing data through sound, beyond the use of this particular software. It is very important to take in mind that its sensorial counterpart, visual exploration, is taught from the beginning of the school stage.

Concerning the last question ("What is the opinion of the participants regarding the multimodal analysis of astrophysical data?"): All participants expressed that this is a novel, promising technique since it allows currently excluded people to be able to explore astronomical data. One of the participants (group C) expressed that they would be able to continue their research ("I think it could be useful for me, to continue my research") ("It allows me to research things that I would have found very difficult before").

As a final reflection, the study demonstrates a marked need to be able to explore their data in an autonomous and effective way. The participants without disabilities showed a marked concern about the possibility of being able to manipulate their data with this new tool. As for the participants with disabilities, they highlighted the possibility of greater autonomy in data analysis that the software allows them (something that so far, no tool allows them), but they are still concerned about the reliability and precision of the tool.

## 5. Conclusion

Understanding a context in digital technology as the circumstances and environment framework present when a person uses a digital tool, the generalisation of this context is very dangerous. Even when this context is catered, many individuals still not being able to accommodate and cope with many digital interfaces. The responses of focus groups participants denote being overwhelmed by digital interactions, feeling minimised, left out, out of voice, feeling inadequate and with the responsibility to lighten a systematically produced heavy air. It was evident during the session, looking at their extra comments and body language that people with disabilities are used to thanks even when the tool is not useful and tend to blame themselves for the mistakes or problems with the tool. Moreover, they don't express any kind of disappointment or negative comment directly, all the recommendations they made were based on a positive comment or even a gratitude.

While it is true that it is challenging to accommodate everyone's needs to create a single static display, it is also very true that it is possible to utilise mechanisms regularly employed to anticipate how humans think and react to use personal data to bombard us with unwanted marketing. These same mechanisms could be used to create a framework that will permit people to come with what they have (their context) and adapt themselves with little or no effort to the display, so people may feel accommodated and empowered to produce or finish a task at their own maximum potential.

Our results reinforce the idea that the HCI should support the user to be successful without suffering fatigue or memory and cognitive overload. During user interface development, assumptions should not be made (for example: the assumption that each person uses the computer, excludes people who do not have this ability; the assumption of everyone being able to interpret the displayed information following the same multi-sensorial clues leaves out people with sensorial biases). The best practice here is to perform high granularity focus group or interviews analysis with potential users and to consider possible user-centred solutions.

People have expressed on several occasions that sonoUno allows them to explore data sets in a new way (albeit in its limited first version). Human beings, by nature, need to explore. If modern computing tools make possible the creation of interfaces that allow people with disabilities autonomy and exploration to work with digital data, why is there hesitancy to develop these on large scales? Why is it so difficult for developers to work in a team with people with disabilities? Functional diverse people need tools to be autonomous, tools that ensures them precision, certainty, effectiveness and efficiency in their work and in equal conditions.

Achieving a universal tool may seem to be utopic, but the mistake is to think small; instead, a more ambitious approach and effort is needed: just as the pieces of a grand jigsaw puzzle fitting together, each assistive technology can form one component of the grand design, all behind the same communication protocol (meaning communication as not only verbal, also between computer programs). In this ideal scenario, people can choose whatever assistive technology and computer program they wish, or fits their needs, and make it work. We may not be so far from being able to realise this; in the same way that we study maths and grammar from childhood, we need to begin by talking about user-centred design and a multi-sensorial approach to data analysis.

## Data Availability

The software, together with the sample data used herein, are available freely in its GitHub repository (https://github.com/sonoUnoTeam/sonoUno). The old version used during the focus group session could be asked to the authors.

All material available on sonoUno can be found at the project web page (https://www.sonouno.org.ar/).

## Acknowledgements

The authors want to thank the support of Southampton Sight (http://southamptonsight.org.uk/), the support from Southampton University Public Engagement with Research Unit that facilitated a stay and all the logistics to perform the Focus Group activities, which was the main topic of this research. The Grant from the CONICET is also appreciated and the IAU-Office for Astronomy Development for facilitating a short stay at the South African Observatory to discuss the first steps for this development. Finally, a special thanks to all the volunteers in this study, their help permitted the authors to concrete the present contribution which will be used to improve next versions of sonoUno-desktop and will be the base for the design of the web interface.

## References

[1] United Nation (1948). Universal Declaration of Human Rights. https://www.un.org/en/about-us/universal-declaration-of-human-rights. Last accessed 5 April 2022.

[2] United Nation (2006). Convention on the Rights of Persons with Disabilities and Optional Protocol [pdf file]. https://www.un.org/disabilities/documents/convention/convoptprot-e.pdf. Last accessed 5 April 2022.

[3] Holloway, C. (2019). Disability interaction (dix) a manifesto. Interactions, 26 (2), 44-49. doi: 10.1145/3310322.

[4] World Health Organization, (2002). The World health report: 2002: Reducing the risks, promoting healthy life. https://apps.who.int/iris/handle/10665/42510. Last accessed 5 April 2022.

[5] Vornholt, K., Villotti, P., Muschalla, B., Bauer, J., Colella, A., Zijlstra, F., Van Ruitenbeek, G., Uitdewilligen, S. and Corbière, M. (2018). Disability and employment–overview and highlights. European journal of work and organizational psychology, 27 (1), 40-55. doi: 10.1080/1359432X.2017.1387536.

[6] Shakespeare, T. (2017). Disability: The Basics (1.ª ed.). London: Routledge.

[7] ISO, (2018). ISO 9241-11: 2018 (en) Ergonomics of human-system interaction - Part 11: Usability: Definitions and concepts. https://www.iso.org/obp/ui/#iso:std:iso:9241:-11:ed-2:v1:en. Last accessed 5 April 2022.

[8] Grabowski, N. A., & Barner, K. E. (1998). Data visualization methods for the blind using force feedback and sonification. Proc. SPIE 3524, Telemanipulator and Telepresence Technologies V (Vol. 3524, pp. 131-139). doi: 10.1117/12.333677.

[9] Mansur, D. L., Blattner, M. M., & Joy, K. I. (1985). Sound graphs: A numerical data analysis method for the blind. Journal of medical systems, 9 (3), 163-174. doi: 10.1007/BF00996201.

[10] Quinton, M., McGregor, I., & Benyon, D. (2016). Sonifying the solar system. In Proceedings of the 22th International Conference on Auditory Display.

[11] Riber, A. G. (2018). Planethesizer: approaching exoplanet sonification. In Proceedings of the 24th International Conference on Auditory Display, 219-226.

[12] Davison, B. K. & Walker, B. N. (2007). Sonification Sandbox reconstruction: Software standard for auditory graphs. In Proceedings of the 13th International Conference on Auditory Display, 509-512.

[13] Díaz-Merced, W. L., Candey, R. M., Brickhouse, N., Schneps, M., Mannone, J. C., Brewster, S. & Kolenberg, K. (2011). Sonification of astronomical data. Proceedings of the International Astronomical Union, 7 (S285), 133-136.

[14] Worrall, D., Bylstra, M., Barrass, S. & Dean, R. (2007). SoniPy: The design of an extendable software framework for sonification research and auditory display. In Proceedings of the 13th International Conference on Auditory Display, 445-452.

[15] Cooke, J., Díaz-Merced, W. L., Foran, G., Hannam, J. & García, B. (2017). Exploring data sonification to enable, enhance, and accelerate the analysis of big, noisy, and multi-dimensional data: workshop 9. Proceedings of the International Astronomical Union, 14 (S339), 251-256.

[16] Casado, J., Díaz-Merced, W. L. & García, B. (2021). Recommendations for accessible human computer interface (HCI) design [pdf file]. https://www.sonoUno.org.ar/wp-content/uploads/sites/9/2021/07/Recommendations-accessible-HCI-design-2021.pdf. Last accessed 6 April 2022.

[17] Casado, J., García, B. & Díaz-Merced, W. L. (2019). Analysis of astronomical data through sonification: reaching more inclusion for visually disable scientists. Proceedings of the International Astronomical Union, (S358), in press.

[18] Bahr, G. S. & Ford, R. A. (2011). How and why pop-ups don't work: Pop-up prompted eye movements, user affect and decision making. Computers in Human Behavior, 27 (2), 776-783. doi: 10.1016/j.chb.2010.10.030.

[19] Massengale, L. R. & Vasquez III, E. (2016). Assessing Accessibility: How Accessible Are Online Courses for Students with Disabilities? Journal of the Scholarship of Teaching and Learning, 16 (1), 69-79.

[20] Malterud, K. (2012). Systematic text condensation: a strategy for qualitative analysis. Scandinavian journal of public health, 40 (8), 795-805. doi: 10.1177/1403494812465030.

[21] Cohen, J. (1960). A coefficient of agreement for nominal scales. Educational and psychological measurement, 20 (1), 37-46. doi: 10.1177/001316446002000104.

[22] Lombard, M., Snyder-Duch, J. & Bracken, C. C. (2002). Content analysis in mass communication: Assessment and reporting of intercoder reliability. Human communication research, 28 (4), 587-604. doi: 10.1111/j.1468-2958.2002.tb00826.x.

[23] Hallgren, K. A. (2012). Computing inter-rater reliability for observational data: an overview and tutorial. Tutorials in quantitative methods for psychology, 8 (1), 23.

[24] Sim, J. and Wright, C. C. (2005). The kappa statistic in reliability studies: use, interpretation, and sample size requirements. Physical therapy & Rehabilitation Journal, 85 (3), 257-268. doi: 10.1093/ptj/85.3.257.

[25] McHugh, M. L. (2012). Interrater reliability: the kappa statistic. Biochemia medica, 22 (3), 276-282.

[26] Supper, A. (2012). The search for the "killer application": Drawing the boundaries around the sonification of scientific data. In The oxford handbook of sound studies. doi: 10.1093/oxfordhb/9780195388947.013.0064.